\documentstyle[12pt
]{article}
\newcommand{\bc}{\begin{center}}
\newcommand{\ec}{\end{center}}
\newcommand{\bd}{\begin{displaymath}}
\newcommand{\ed}{\end{displaymath}}
\newcommand{\be}{\begin{equation}}
\newcommand{\ee}{\end{equation}}
\newcommand{\ba}{\begin{array}}
\newcommand{\ea}{\end{array}}
\newcommand{\bt}{\begin{tabular}}
\newcommand{\et}{\end{tabular}}
\newcommand{\un}{\underline}

\begin{document}

~
\vspace{4cm}
\begin{center}
{\LARGE\bf{New approach to the non-perturbative physics}}\\[2mm]
{\bf V.G.Ksenzov}\\ [2mm]
\bf{Institute of Theoretical and Experimental \\Physics}
\end{center}

\vspace{1cm}
\begin{abstract}
It was shown that the non-perturbative properties of the vacuum are
described by the quantum fluctuations around the classical background
with zero canonical momentum. The vacuum state has been built and checked
in the framework of the sigma models in two dimensions.
\end{abstract}

\vspace{9cm}

\un{~~~~~~~~~~~~~~~~~~~~~~~~~~~~~~~~~~}

e-mail: ksenzov@vitep5.itep.ru

\newpage

In particle physics complicated problems of non-abelian gauge
theories are rathe often investigated in  simple models.
Doing so, people hope that these studies will lead to  an
understanding of the qualitative behaviour of the actual dynamics.
One of such simple models is the two dimensional non-linear sigma
model with fields transforming under the vector representation of the
group $O(N)$. The model provides a possibility to get a set of
non-trivial results of non-perturbative physics [1,2] which are the
keys to the calculation of the vacuum condensates.

Now there are a few different approaches to the calculation of the
vacuum condensates in simple models. One of the approaches is to use
effective potentials. This method allows one to obtain  the vacuum
expectation value (v.e.v.) of different operators and to judge the
availability of the non-perturbative effects [1], but
non-perturbative dynamics cannot be investigated via using effective
potentials. There are papers in which the vacuum structures of the
models are studied using a variational technique with pairing ansatz
for the trial vacuum states [3-5]. Use of the generating
functional for the Green function  has made it possible to clear up a
role of the non-perturbative effects in producing the
non-perturbative part of the vacuum energy density in sigma models
[6,7].

In the present paper our goal is to build the vacuum state in the
framework of the two-dimensional sigma models in the large $N$ limit.
It will be  shown that besides the classical fields there are
special quantum fields defining the non-perturbative properties of
the vacuum state. In non-trivial cases these quantum fields and a
combination of classical fields may generate scaling independent v.e.v.
usually being got by effective potential [7].
There are a few properties of the vacuum state that we 
 know should be satisfied.Therefore using the vacuum state we should
obtain v.e.v. of the energy-momentum tensor and  show that the 
momentum of the vacuum is zero and,in addition,the vacuum
state is Lorentz invariant.    Although our
results are derived in the special case of sigma models it can  be proposed
that the physical idea of describing vacuum state  may be of general
interest.

The physical idea of decomposing quantum fields is based on assumption
that the non-perturbative properties of the vacuum are defined by the
quantum fields with conjugated canonical momentum which must be zero.
In this case such quantum fields are not operators but are $c$-number
 functions.

Our assumption can be better explained in terms of a model having  gauge
fields $F^a_{\mu\nu}$ and isovector fields $\phi^a$ ($a$ is a
group index) with the group O(3) [8,9]. The Hamiltonian of the model is  
defined as
$$
H=\frac{1}{2}\int
d^2x(B^2+E^2)+(D_{\mu}\phi^a)^2+\lambda^2(\phi^2-a^2)^2,
$$
where $E^k=F^{0k}$, $B_k=-\frac{1}{2}\epsilon_{ijk}F^{ij},
D_{\mu}\phi^a=\partial_{\mu}\phi^a+g\epsilon^{abc}A^b_{\mu}\phi^c$.
It is well known that finite energy requires that the contribution of each
positive definite term to the energy density is final and vanishes at
large distances. The finite energy condition is satisfied at
large distances as long as
$$
A_{\mu}=0,\qquad \phi_0=const ,
$$
or
$$
A_{\mu}=U\partial_{\mu}U^{-1},\qquad D_{\mu}\phi=0,\qquad
\phi{(x)}=U\phi=exp\;\alpha^a(x)t^a\phi_0,
$$
where $t^a$ is a generator of the group $O(3)$, $\phi_0$ is a fixed vector
minimizing the potential. In this case the phase factor
$\alpha^a(x)$ is the integral of the conserved symmetry
current $I^a_{\mu}=\epsilon^{abc}\partial_{\mu}\phi^b\phi^c$ 
of the $\phi^a$--fields. Really, differentiating
$$
\phi(x)=exp\alpha^a(x)t^a\phi_0,
$$
we get
$$
\partial_{\mu}\phi^a(x)=\partial_{\mu}\alpha^b(x)\epsilon^{abc}\phi^c(x),
$$
then, multiplying  the equation, by $\epsilon^{ka\ell}\phi^{\ell}$ we
obtain
$$
I^k_{\mu}=\partial_{\mu}\alpha^k(x)\phi^2-(\partial_{\mu}\alpha^a(x)\phi^a
(x))\phi^k.
$$
Here  we chose $\phi^2=a^2$. After all, the solution of the equation
is
$$
\alpha^a(x)=\frac{1}{a^2}(\int\limits^xdz^{\mu}I^a_{\mu}(z)+\int\limits
^xdz^{\mu}n_{\mu}(z)\phi^a(z)),
$$
where $n_{\mu}$ is an arbitrary function. In addition we imply that
the $A_{\mu}$ -- fields are subject to the gauge fixed condition
$\partial_{\mu}A_{\mu}=0$, therefore $n_{\mu}$ must be zero.

To obtain  quantum field theory it is necessary to require that
fields and canonical momenta $\Pi^a_0$ are operators satisfying some
commutation relations. In the model under consideration the canonical momenta
$\Pi^a_{\mu}$ are defined as $\Pi^a_{\mu}=D_{\mu}\phi^a(x)$ and must
tend to zero at large distances. However, it can be conceived that
the fields take the vacuum values at infinity therefore it may be
thought that at any point x the conditions $\Pi^a_{\mu}=0$ ensure the
vacuum values of the $c$-number scalar fields, i.e. we imply that the
scalar vacuum fields are covariant constants with the connection
$I^a_{\mu}(x)$. Note that the conditions $\partial_{\mu}\phi^a=0$,
$A_{\mu}=0$ or $D_{\mu}\phi^a=0$ satisfy the requirements
$\Pi^a_{\mu}=0$. As it can be  seen later, the first of them defines the
perturbative vacuum  while second requirement gives the
non-perturbative vacuum unless $\phi^a$ is zero.

In the rest of this paper it will be discussed how to build the vacuum
states in the framework of sigma models in the large $N$ limit to
check this hypothesis.

The Lagrangian of the $O(N)$ sigma models is taken in Minkowski
space-time in the form
$$
L=\frac{1}{2}\left\{\frac{N}{f}(\partial_{\mu}\sigma^a)^2+\frac{\alpha(x)}
{\sqrt{N}}(\sigma^a\sigma^a-1)\right\},\qquad a=1,2...N.
$$
The factor $N^{-1/2}$ under the Lagrange multiplier field $\alpha(x)$
is written for the sake of convenience.
In order to set off the non-perturbative contributions in the vacuum
states one decomposes $\sigma^a(x)$ and $\alpha(x)$ as
$$
\sigma^a(x)=c^a(x)+q^a(x),
$$
$$
\alpha(x)=\alpha_c(x)+\alpha_{\mu}(x),
$$
where $c^a(x)$ and $\alpha_c(x)$ are classical fields, while $q^a(x)$
and $\alpha_{qu}(x)$ describe small fluctuations around classical
background.

The dynamics of classical and quantum fields is described by the
Lagrangians [1]
$$
L_c=\frac{1}{2}\left\{\frac{N}{f}(\partial_{\mu}c^a)^2+\frac{\alpha_c(x)}
{\sqrt{N}}(c^ac^a-1)\right\},
$$
$$
\L_{qu}=\frac{1}{2}\left\{\frac{N}{f}(\partial_{\mu}q^a)^2+\frac{\alpha_c}
{\sqrt{N}}q^aq^a+\frac{\alpha_{qu}}{\sqrt{N}}(c^aq^a+q^aq^a)\right\}.
$$
The classical fields satisfy the equations of motion
\be
-\frac{N}{f}(\partial_{\mu}c^a)^2=\frac{\alpha_c(x)}{\sqrt{N}},
\ee
$$
c^a(x)c^a(x)=1.
$$
We make the simplifying assumption that the quantum fields satisfy
the equations of motion in the linear form with respect to the quantum
fluctuations. We have
\be
\frac{N}{f}\partial^2q^a=\frac{\alpha_c}{\sqrt{N}}q^a+\frac{\alpha_{qu}c^a}
{\sqrt{N}},
\ee
\be
c^a{(x)}q^a{(x)}=0.
\ee
In the rest of the paper it will be discussed the quantum fields which
are subject to the constraint (3). The solution of eq.(3) is obtained
in ref [7]. If at any point, say $x=0$, $q^a(0)$ is chosen in such a
way that $c^a(0)q^a(0)=0$ then $q^a(x)$ can be found at any other
point $x\not= 0$ so that $c^a(x)q^a(x)=0$ due to $c^a(x)$ and
$q^a(x)$ are constants up to a phase factor. Indeed, differentiating
(3) and using the identity [6]
\be
\partial_{\mu}c^a=-I^{ab}_{\mu}c^b,
\ee
here
$I^{ab}_{\mu}(x)=c^a(x)\partial_{\mu}c^b(x)-\partial_{\mu}c^a(x)c^b(x)$, for
arbitrary $c^a(x)$ we get
\be
\partial_{\mu}q^a(x)+I^{ab}_{\mu}(x)q^b(x)=0.
\ee
Let $u^a(x)$ be a special form of the solutions of eq.(5). Then they
can be written as
\be
u^a(x)=P\;exp\left(-\int\limits^x_0dz^{\mu}I_{\mu}(z)\right)_{ab}q^b(0).
\ee
The formal solution of eq.(4) is written in the form
\be
c^a(x)=P\;exp\left(-\int\limits^x_0dz^{\mu}I_{\mu}(z)\right)_{ab}c^b(0).
\ee
One can check that $c^a(x)q^a(x)=0$, $c^a(x)c^a(x)=1$ and
$u^a(x)u^a(x)=q^a(0)q^a(0)=const\not=0$, i.e. $u^a(x)$ fields are
also subject to the second class constraint.

Now we can discuss the vacuum structure of sigma model. 
Let us proceed to the construction of the perturbative vacuum
structure for a pedagogical intent.

Naively the canonical momenta of the fields which are subject to the
constraint (3) are $\Pi^a_{\mu}=\partial_{\mu}q^a$. In accordance
with the assumption, $c$-number vacuum fields are the solutions of
the equations $\Pi^a_{\mu}=\partial_{\mu}q^a=0$, i.e. $q^a(x)=v^a$
$(v^2\ll1)$ are constants as well as $c$'s . Then from eq.(1)
we get $\alpha_c=0$, i.e. the case corresponds to the description of
the  perturbative vacuum [1]. Note that the quantum
Lagrangian is not invariant under the substitution $q^a\to q^a+v^a$
but the Lagrangian without the interaction terms is.  The
operator which carries out the transformation $q^a\to v^a$ can be
found from the free Lagrangian by using the Noether theorem.  However
it is well known that this is the operator of the canonical momentum.  

The vacuum of the free theory may be described as the
eigenfunction of the annihilation operator $a^k_0$ with zero
eigenvalue of the momentum[10] $$ a^k_0|v>=v^k|v>, $$ where $|v>=exp(v^a\int
dx\partial_0q^a)=exp(v^b(a^b_0-a^+_0))|0>$, and $a^b_0|0>=0$.

We can
build the non-perturbative vacuum state by separating the quantum
fields into two categories: the fields $u^a$ satisfy (5) and all the rest
of the fields $q^a(x)$. Firstly, we argue that the canonical momenta
conjugate to the fields $u^a$ are equal to zero. To do this the
Lagrangian of the quantum fields   should be written as
$$ L_{qu}=L_{qu}(u)+L_{qu}(a), $$ where $$
L_{qu}(u)=\frac{1}{2}\frac{N}{f}\{(\partial_{\mu}u^a)^2-(I^{ab}_{\mu}u^b)
^2-(I^a_{\mu}u^a)^2\},
$$
$$
L_{qu}(q)=\frac{1}{2}\left\{\frac{N}{f}(\partial_{\mu}q^a)^2+\frac{\alpha_c}
{\sqrt{N}}q^aq^a\right\}.
$$
To obtain $L_{qn}(u)$ use was made of the identity
$$
\frac{\alpha_c}{\sqrt{N}}u^au^a=-\frac{N}{f}(\partial_{\mu}c^u)^2u^au^a
=-\frac{N}{f}(I^a_{\mu}(x))^2u^2,
$$
where $I^a_{\mu}=\frac{1}{2}f^{abc}I^{bc}_{\mu}$, $f^{abc}$ is
antisymmetric tensor satisfying the relation
$$
f^{abc}f^{ake}=(\delta^{bk}\delta^{ce}-\delta^{bc}\delta^{bc}).
$$
In this case we have
$$
(I^{ab}_{\mu}u^b)^2=(I^a_{\mu})^2u^2-(I^a_{\mu}u^a)^2.
$$
Making use of (6) we obtain
\be
L_{qu}(u)=\frac{N}{2f}[\partial_{\mu}u^a(\partial_{\mu}u^a+I^
{ab}_{\mu}u^b)-(I^a_{\mu}u^a)^2]
\ee
and the canonical momentum of $u^a$ is
\be
\Pi^a_{\mu}=\frac{\delta
L}{\delta(\partial_{\mu}u^a)}=\partial_{\mu}u^a+I^{ab}_{\mu}u^b=0.
\ee
Therefore the fields $u^a$ define the non-perturbative vacuum state.

Let us find the operator of the transformation $q^a(x)\to u^a(x)$,
such that $D^{-1}q^a(x)D=u^a(x)$ and $D=exp G$. The free
Lagrangian of the fields $u^a$ is
\be
L=\frac{1}{2}\frac{N}{f}(\partial_{\mu}u^a(\partial_{\mu}u^a+I^{ab}_
{\mu}u^b)).
\ee
Then the variation of the Lagrangian without using the equations of
motion is
$$
\delta
L^{(1)}=\frac{1}{2}\frac{N}{f}\partial_{\mu}(u^aI^{(ab)}_{\mu}q^b)
$$
and with using the equations of motion is
$$
\delta L^{(2)}=\frac{N}{2f}\partial_{\mu}(u^a\partial_{\mu}q^a).
$$
Note that the equations of motion are obtained from the free Lagrangian
(10) and reduce to eq. (5). Requiring that the difference of two
variations is equal to zero we get
$$
\delta L^{(1)}-\delta
L^{(2)}=\partial_{\mu}I_{\mu}=\partial_{\mu}[u^a(\partial_{\mu}q^a-I^{ab}
_{\mu}q^b)]=0,
$$
i.e. there is the conservation current for the free Lagrangian
\be
I_{\mu}=u^a(x)(\partial_{\mu}q^a(x)-I^{ab}_{\mu}q^b(x)).
\ee
Then the charge $G$ is defined as
\be
G=\int dx u^a(x)(\partial_0q^a-I^{ab}_0q^b).
\ee
The non-perturbative vacuum state may be defined as
\be
D^{-1}|0>=|u>.
\ee
It is known [1,7] that the non-perturbative part of the vacuum energy
density $\varepsilon^n_{vac}$ is written as
\be
\varepsilon^n_{vac}=\frac{1}{2}<0|\theta^R_{\mu\mu}|0>=-\frac{\beta(f)}
{4f^2}N\;lim
{\hspace{-0.78cm}\raisebox{-0.3cm}{$\scriptstyle
\Delta\to0$}\;\,}
<0|\partial_{\mu}q^a(x-\Delta)\partial_{\mu}q^a(x+\Delta)|0>.
\ee
Here the quantity $\theta^R_{\mu\mu}$ has been regularized by
separating the quantum fields at different points. Although the
energy of the vacuum $\varepsilon^n_{vac}$ has been calculated by means of the
generating functional in [7] our purpose is to check the
non-perturbative vacuum state (14) by describing v.e.v. of the
operators $\theta^R_{\mu\mu}$, $P_{\mu}$ and Lorentz transformation
$M^{ik}$. To do this we should remind that now the quantum fields
$q^a$ in (14) have to be considered as the perturbative part of the
quantum fluctuations, i.e. the fluctuations which satisfy eq. (6) are
omitted from the $q^a(x)$ fields and the perturbative state $|0>$
must be replaced by the non-perturbative vacuum state $|u>$.

Then (14) may be rewritten as
\be
\varepsilon^n_{vac}=-\frac{\beta(t)}{4f^2}N<u|\partial_{\mu}q^
a(x-\Delta)\partial_{\mu}q^a(x+\Delta)|u>.
\ee
Substituting (13) into (15) we obtain the vacuum energy
\be
G^n_{vac}=\frac{1}{2}<0|\theta^R_{\mu\mu}|0>-\frac{1}{2}<0|[G\theta
^R_{\mu\mu}]|0>+\frac{1}{4}<0|[G[G\theta^R_{\mu\mu}]]|0>+\ldots
\ee
Here the first term is the perturbative contribution which vanishes,
as it is shown in [1], the second term is zero due to
$<0|\partial_{\mu}q^a|0>=0$.

Then in the Euclidian space  we get 
\be
\varepsilon^n_{vac}=\frac{\beta(t)}{4f^2}(\partial_{\mu}c^a(x))^2<0|
u^a(x-a)u^a(x+\Delta)|0>.
\ee
As far as (17) has been calculated in [7] and it can be written
as
$$
\varepsilon_{vac}=\frac{N}{8\pi}\frac{1}{f}M^2exp(-4\pi/f)u^2(0).
$$
The quantity $u^2(0)$ is a free parameter, which can be chosen in
such a way that $u^2(0)=f$ and we obtain the known result [1]. Note
that the remaining commutation terms equal zero in (16).

In the similar manner one can obtain that $<u|P|u>=0$, where $P=\int
dx\partial_0q^a\partial_1q^a$

The operator of the Lorentz transformation is defined as
$$
M^{01}=\int dx(x\theta^{00}-t\theta^{01}),
$$
and, clearly,
$$
\omega^{01}<u|M^{01}|u>=\omega^{01}(\int dx x<u|\theta^{00}|u>-t\int
dx<u|\theta^{01}|u>)=0,
$$
as $<u|\theta_{00}|u>=const$ and $<u|\theta_{01}|u>=0$.

Now it remains to show how in the perturbative theory the results emerge
which are related to the non-perturbative effects. We recall that the
non-perturbative part of the vacuum energy density has been obtained
within the perturbative calculations [1]. The matter is that in such a
situation the naive definition of the conjugate canonical
momentum $\pi^a_{\mu}=\partial_{\mu}q^a\not=0$ contains any $q^a(x)$
and, consequently, $u^a(x)$ due to $\partial_{\mu}u^a\not=0$.
Therefore $u^a$ fields appear in calculations similar to the perturbative
fields but the commutation relations for $u^a$ are different from
those for $q^a$. The $u^a$ fields are subject to the second class
constraint because $u^a(x)u^a(x)=const\not=0$ (6). Indeed from (3)
and (5) we can obtain the following commutation relations for the
$u$'s and $\pi$'s:
$$
[u^a, u^b]=0\qquad [\pi^au^b]=\delta^{ab}-c^a(x)c^b(x)
$$
\be
[\pi^a, \pi^b]=-I^{ab}_0(x)
\ee
The commutation relations (18) may be obtained in a more simple way
if  the commutation relations  for the sigma fields [5] are
used where the sigma fields are decomposed into classical and quantum fields.

It is precisely due to this reason that the perturbative vacuum of
the theory which is annihilated by all the a's operators may by used
to calculate the non-perturbative effects but in terms of the quantum
fluctuations being subject to the second class constraint.

In this paper we have performed the physical idea which is based on
the assumption that the non-perturbative properties of the vacuum are
described by quantum fluctuations around the classical background
with the conjugate canonical momentum is equal to zero. The assumption
has been checked by calculating the vacuum energy density and the
momentum of the vacuum. 
The Lorentz invariance of the vacuum state was proved in the framework
of sigma models. It was discussed that the non-perturbative
effects can be described by the quantum fields which are subject to
the second class constraint.

The author wishes to thank V.A.Novikov, A.D.Mironov and A.E.Kudryavtsev
for useful discussions.

The paper was partially supported by the
Russian Foundation of Fundamental Research (grant ü 95-02-04681-A).

\end{document}